\documentstyle[prb,aps,twoside,twocolumn,epsf]{revtex}
\begin{document}

\title{\boldmath
Computer simulation of low-energy excitations in
amorphous silicon with voids
}
\author{
 Serge M. Nakhmanson and
 D. A. Drabold\\
 {\small\it
  Department of Physics and Astronomy, Ohio University,
  Athens, Ohio 45701-2979\\ 
 }
}
\date{\today}
\maketitle

\begin{abstract}
\noindent
We use empirical molecular dynamics technique to
study the low-energy vibrations in a large 4096 atom model
for pure amorphous silicon and a set of models with voids
of different size based on it. Numerical vibrational eigenvalues 
and eigenvectors for our models are obtained by exact 
diagonalization of their dynamical matrices. 

Our calculations show that localized low-energy vibrational
excitations of rather complex structure are present in
amorphous silicon models with voids. According to their
spatial localization patterns we make an attempt to 
classify these excitations as modes associated with the void
and  ``mixed'' modes associated with the interaction of the
void with strained regions of silicon network.

\end{abstract}
\pacs{63.50.+x, 63.20.Pw, 61.43.Dq}

\section {Introduction}
Using {\it ab initio} molecular dynamics (MD) in a small
216 atom model for amorphous silicon
we have recently demonstrated\cite{ai_bubbles} that a presence of 
spherical void type defect in  a-Si network results 
in localized  low energy excitations. To gain a better 
understanding of the behavior of these vibrational 
modes  and verify our previous results
we perform extensive computer MD 
simulations for large 4096 atom model for a-Si and models
with voids constructed from it.
Thanks to the large size of the supercell for this model
(approximately 43~\AA) we are able to track the ``void
size dependent properties'' of the system by building a set of 
models with bigger and bigger voids, but of course at the
expense of losing all the merits of {\it ab initio} approach,
which makes the calculations way too slow for the models
containing thousands of atoms.

In our current investigation we employ an 
en\-vi\-ron\-ment-de\-pen\-dent interatomic potential (EDIP) for
amorphous silicon
developed by Bazant and Kaxiras\cite{Bazant1}
which enables us to dramatically improve the speed of calculations
in comparison to {\it ab initio} technique. Due to the fact that
this potential is (i) relatively new and (ii) has been fitted to simulate
only the bulk properties of amorphous silicon\cite{Bazant2} 
(and not surfaces --- like we get when
making a void in our model) it is also interesting to test it's
accuracy for calculations of vibrational properties of models for
pure a-Si and a-Si with voids, especially on small ones --- to verify
that EDIP can reproduce (at least, qualitatively) the features we
see with {\it ab initio} method.

In Sec.\ II of this paper we describe the construction routine for 
the models we study and the calculation scheme for analyzing
their vibrational properties. In Sec.\ III we discuss the properties
of vibrational excitations
obtained for 216 atom (for verification purposes) and 4096 atom 
models and their derivatives with voids, 
including their spatial localization and dependence on the size of
the void. 
We conclude in Sec.\ IV by giving a summary of our results
on computer simulation of low-energy excitations in
amorphous silicon.

\section {Models, their construction and calculation scheme}
The model construction routine we make use of closely resembles
the one that is described in our previous paper\cite{ai_bubbles}
with a single major difference: in it's current state EDIP can
work only with one atom type --- silicon, which makes it impossible
to simulate hydrogenated a-Si. 

In everything else our model constructing routine works as follows:
we employ 4096 atom model for pure a-Si made by Djordjevic,
Thorpe and Wooten\cite{WWWupd} (refered to as DTW in what
follows) with the help of Wooten, Winer
and Weaire bond switching algorithm\cite{WWW} as a ``base'' for
building a family of models with voids.
The model is relaxed with our MD code with EDIP until the forces
on every atom in the system are smaller than $0.01$~eV/\AA.
Then an arbitrary atom\cite{atom}
is chosen and all the atoms (including the chosen one) within 
certain radius from it are removed to make a void. 
After this procedure the model with a void is relaxed once again to obtain
it's equilibrium configuration, so that the harmonic approximation
for the total energy of the system is appropriate. 
Finally we compute the dynamical matrix of the system by displacing
every atom in the cell by $0.03$~\AA\  in three orthogonal directions
and calculating the originating EDIP forces on all the atoms in the 
system. Due to the fact that dynamical matrix for a system containing 
thousands of
atoms is huge we have to extensively exploit it's sparse
nature discarding terms smaller than 
\hbox{$10^{-4} \hbox{eV} \hbox{\AA}^{-2} {a.u.m.}^{-1}$},
which in our opinion is a good compromise between the accuracy 
of the calculation and compactness of the output.
Once the sparse dynamical matrix for the system is obtained we use a separate
computer program to calculate it's eigenvalues and eigenvectors together with
their inverse participation ratios (IPR). Again, for the sake of compactness,
only low-energy (less than \hbox{$200~cm^{-1}$}) eigenvectors with
significantly high IPRs --- and this is the point of our main interest --- are 
written out.

We make use of gaus\-si\-an broade\-ned 
representation for \hbox{$\sum_i \delta(E-E_i)$}, where
\hbox{$E_i, i = 1 \ldots N$} are the eigenvalues of the dynamical
matrix to plot the graphs for vibrational density of states (VDOS) for
our system. The width of broadening is \hbox{$20~cm^{-1}$} for the
``full scale'' graphs and \hbox{$0.1~cm^{-1}$} for the close-ups of the
low-energy region.

We finish our investigation by creating color or gray scale vibrational
activity maps for the ``low-energy, high IPR'' modes in exactly the
same way that has been already described in Sec.\ II of Ref.~1.

\section {Discussion of results}

\begin{figure*}[h]

\epsfxsize=7cm
\moveright 1cm \vbox{\epsfbox{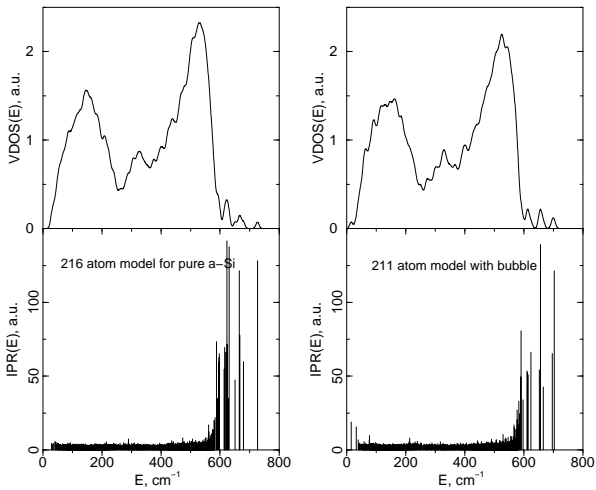}}

\vskip -1cm

\caption{ 
\label{si216}
VDOS and IPR for ``pure'' 216 atom DTW model
for a-Si (left set of panels) and 211 atom
model with void (right set of panels).}
\end{figure*}

We begin this section with presenting some testing results
for small 216 atom DTW model and 211 atom model with a void
constructed from it by removing a single atom from the network
together with it's four nearest neighbors. The results of our
VDOS and IPR calculations are shown in Fig.~\ref{si216}.
It's easy to notice that the model with void has a localized
state in the low-energy gap which complies with our previous
findings\cite{ai_bubbles} (we neglect all of the hydrogen motion
in our a-Si:H model in Ref.~1 while making this comparison
because hydrogen atoms do not ``participate'' in vibrational
excitations of such low energy). It also means that despite the fact
that EDIP, comparing to the {\it ab initio} calculation, 
gives us different low-energy gap edge,
it is capable of reproducing
the localized low-energy excitations. A ``vibrational
activity'' colormap for the excitation we see in 211 atom model
with void (not shown here) also appeares to be in good agreement with our
previous results. 

\vskip -1cm

\begin{figure*}[h]

\epsfxsize=7cm
\moveright 1cm \vbox{\epsfbox{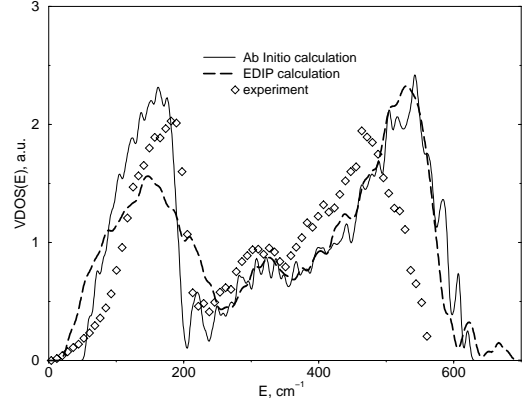}}

\vskip -1.5cm

\caption{ 
\label{vdos216}
A comparison of VDOS for 216 atom DTW model
calculated with {\it ab initio} 
and EDIP. }
\end{figure*}

The graph with a comparison of VDOS calculated for 216 DTW model
with {\it ab initio} and EDIP is presented in Fig.~\ref{vdos216}. 
The experimental results are taken from Kamitakahara 
{\it et al.\cite{Kamitakahara}} In our opinion this figure 
demonstrates that, at least for simulations of {\it vibrational} 
properties of a-Si, EDIP exhibits the same behavior as the
well known Stillinger-Weber potential\cite{Feldman1}.

\vskip -1.5cm

\begin{figure*}[h]

\epsfxsize=7cm
\epsfysize=10cm
\moveright 1cm \vbox{\epsfbox{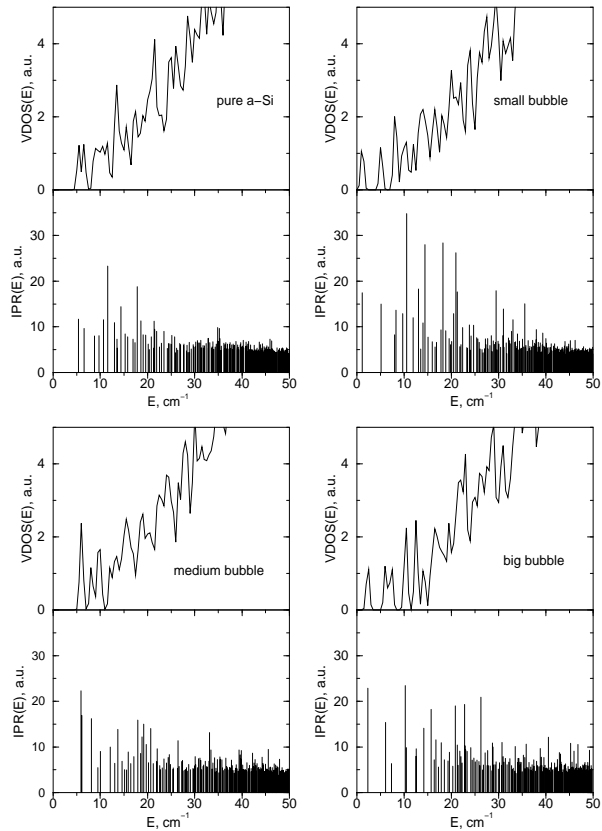}}

\vskip 1cm

\caption{
\label{big_four}
VDOS and IPR low-energy region snapshots 
for ``pure'' 4096 atom DTW model
(upper left set of panels), 4091 atom ``small bubble''
model (upper right), 4069 atom ``medium bubble''
model (lower left) and 4008 atom ``big bubble''
model (lower right).}
\end{figure*}

Now we are going to discuss the large 4096 atom DTW model and the
family of models with voids based on it. Three models with voids
of different size have been built: 4091 atom model with a ``small
bubble'' (analogous to 211 atom model derived from 216) --- a void
of approximately 5~\AA\ in diameter, 4069 atom ``medium bubble''
model with 10~\AA\ void and 4008 atom ``big bubble'' model
with 15~\AA\ void. The low-energy region VDOS and IPR close-ups for 
all the four models are shown in Fig.~\ref{big_four}. We do not
present ``full scale'' VDOS graphs here because (i) in that scale they
all look indistinguishable from each other and (ii) generally, VDOS
constructed for 4096 atom model doesn't provide much more information 
than the one of 216 atom model, shown in Fig.~\ref{si216}.

From Fig.~\ref{big_four} one can see that the situation with the large
models is more complicated than in the case of the family of 216 atom models.
First of all, it turns out that ``pure'' 4096 atom model has two localized
low-energy modes\cite{Feldman2} that are most probably associated with strained
regions of silicon network. Consequently, we receive modes of two
major types in our void models: those associated directly with voids and
modes produced by mixing of the former ones with localized excitations
of the ``pure'' model. The modes, associated with voids, exhibit the same
kind of localization properties that we have reported earlier\cite{ai_bubbles},
localizing to the side of the void, while the modes, produced by mixing,
either localize on approximately the same strained regions of the cell 
as in ``pure'' model
(although they are not exact copies of the modes observed in ``pure'' model,
which we attribute to the fact that quenching of the model with void
results in different system geometry even in the regions away from the void)
or form strings between the strained regions and the void.

\vskip 1cm

Curiously enough our data shows that for different models with voids modes
of different types dominate in the low-energy region! In the ``small bubble'' 
model a mode with the
highest IPR at \hbox{10.58~$cm^{-1}$} is of ``void'' type, but the succeeding
three modes with high IPR at 14.43, 18.25 and \hbox{20.97~$cm^{-1}$}
are of strongly pronounced ``mixed'' type. In the ``medium bubble'' model,
to the contrary, all three low-energy localized modes at 5.89, 6.12
and \hbox{8.13~$cm^{-1}$} are of ``mixed'' type. The mode with strong
``void'' type behavior is also present but it is shifted to 
\hbox{17.97~$cm^{-1}$}. Finally in the ``big bubble'' model
modes at 2.34 and \hbox{6.10~$cm^{-1}$} are of ``void'' type and all the
others, including a strongly localized mode at \hbox{10.28~$cm^{-1}$},
exhibit ``mixed'' type behavior.

For the sake of compactness we haven't included any of the colormaps for
the 4096 atom family of models in this paper. However a colormap set for
particularly interesting vibrational excitations in these models is
available for Internet download\cite{download}.

\section {Conclusions}
We have studied low-energy vibrational excitations in 216 and 4096 atom 
DTW models
for amorphous silicon and the families of models with voids based on them,
employing the new Bazant-Kaxiras en\-vi\-ron\-ment-de\-pen\-dent 
interatomic 
potential. Good qualitative agreement between the data obtained 
for 216 DTW model and it's 
derivative with a void and the results of our previous {\it ab initio} 
calculations for the same model
leads us to believe that EDIP can be used for probing localized 
low-energy modes in a-Si models. Our investigation of vibrational properties
of 4096 atom model and it's void derivatives shows that the latter
posses a complex spectrum of localized low-energy excitations
that we can roughly divide into two groups --- ``void'' and ``mixed'' 
type modes --- according to their localization patterns.
We have also found out that there is no simple dependence between
the size of the void and the energy of it's ``void'' type mode. It seems
that not only the size of the void but also the positions of the void and
strained regions of the network with respect to each other
and peculiarities of the network geometry play a very
important role in understanding why the modes of a certain localization
nature (i.e. ``void'' type or ``mixed'' type) dominate at certain energy
intervals.

\section*{Acknowledgments}
SMN thanks Dr.\ M. Z. Bazant for many helpful discussions and
Dr.\ W. A. Kamitakahara for providing numerical results of 
VDOS measurements for a-Si. 
This work was supported by NSF
under grant number DMR 96-18789 and DURIP N00014-97-1-0315.


\begin{references}
\bibitem{atom}
For convenience, we usually pick atoms positioned close to the center of 
the supercell.
\end{references}
\end{document}